\documentstyle[11pt]{article}
\def \beq{\begin{equation}}
\def \eeq{\end{equation}}
\def\be{ \begin{displaymath} }
\def\ee{ \end{displaymath} }
\def\ben{ \begin{equation} }
\def\een{ \end{equation} }
\def\bea{ \begin{eqnarray} }
\def\eea{ \end{eqnarray} }

\def\ko{ {\rm K}^ 0 }
\def\kob{ \overline{\rm K}{}^0 }
\def\kak{ {\rm K}^ 0 - \overline{\rm K}{}^0 }

\def\kos{ | {\rm K^ 0} \rangle }
\def\kobs{ | {\rm \overline{K}{}^0} \rangle }

\def \bk{\overline K^0}
\def \k{K^0}
\def \h{{\cal H}}

\newcommand{\eu}{\mbox{{\boldmath $e_1$}}}
\newcommand{\ed}{\mbox{{\boldmath $e_2$}}}

\def\kk{ | {\rm K^ 0} \rangle }
\def\kbk{ | {\rm \overline{K}{}^0} \rangle }
\def\koo{ | {\rm K^ 0}(0) \rangle }
\def\bkoo{ | {\rm \overline{K}{}^0}(0) \rangle }
\def\kkl{|K_L^0>}

\def\kks{|K_S^0>}

\def\kku{|K_1^0>}

\def\kkd{|K_2^0>}

\def\kst{|K_S(t)>}
\def\kso{|K_S(0)>}
\def\klt{|K_L(t)>}
\def\klo{|K_L(0)>}
\def\kot{|K^0(t)>}
\def\bkot{|{\overline K}^0(t)>}

\def\eas{e^{-\alpha_S t}}
\def\eal{e^{-\alpha_L t}}
\def\fpt{f_{+}(t)}
\def\fmt{f_{-}(t)}

\def \cp {\par \noindent }
 
\def \kok {K^0- \overline {K^0} }

\def \ee {${\epsilon ' \over \epsilon} $ }

\def \ko {K^0 }
\def \kbo {\overline{K^0} }

\def\ket#1{\vert #1 \rangle }

\def\Ds {D \kern-2.2ex /} 
\def\dw {W \kern-2.2ex /} 
\def\dz {Z \kern-2.2ex /}
\def\dk {k \kern-1.2ex /}
\newcommand{\lsim}{\mathrel{\raise.5ex\hbox{$<$}\kern-.75em\raise
                                             -.5ex\hbox{$\sim$}}}
\newcommand{\gsim}{\mathrel{\raise.5ex\hbox{$>$}\kern-.75em\raise
                                             -.5ex\hbox{$\sim$}}}
\newskip\humongous \humongous=0pt plus 1000pt minus 1000pt
\def\caja{\mathsurround=0pt}
\def\eqalign#1{\,\vcenter{\openup1\jot \caja
        \ialign{\strut \hfil$\displaystyle{##}$&$
        \displaystyle{{}##}$\hfil\crcr#1\crcr}}\,}
\newif\ifdtup

\catcode`\@=11
\def\section{\@startsection {section}{1}{0pt}{-3.5ex plus -1ex minus
 -.2ex}{2.3ex plus .2ex}{\raggedright\large\bf}}
\catcode`\@=12
\catcode`\@=11
\def\eqnarray{\stepcounter{equation}\let\@currentlabel=\theequation
\global\@eqnswtrue
\global\@eqcnt\z@\tabskip\@centering\let\\=\@eqncr
\gdef\@@fix{}\def\eqno##1{\gdef\@@fix{##1}}%
$$\halign to \displaywidth\bgroup\@eqnsel\hskip\@centering
  $\displaystyle\tabskip\z@{##}$&\global\@eqcnt\@ne
  \hskip 2\arraycolsep \hfil${##}$\hfil
  &\global\@eqcnt\tw@ \hskip 2\arraycolsep $\displaystyle\tabskip\z@{##}$\hfil
   \tabskip\@centering&\llap{##}\tabskip\z@\cr}

\def\@@eqncr{\let\@tempa\relax
    \ifcase\@eqcnt \def\@tempa{& & &}\or \def\@tempa{& &}
      \else \def\@tempa{&}\fi
     \@tempa \if@eqnsw\@eqnnum\stepcounter{equation}\else\@@fix\gdef\@@fix{}\fi
     \global\@eqnswtrue\global\@eqcnt\z@\cr}
\newcommand{\npb}[3]{ Nucl.~Phys.\             {\bf B#1}   (#2)   #3}

\def \ite{{\it et al.}}

\begin{document}
\renewcommand{\thetable}{\Roman{table}}

\setlength{\baselineskip}{0.8 cm}

\rightline{UNIBAS--MATH 6/96}
\vspace{2.5cm}

\begin{center}
{\Large \bf THE CLASSICAL ANALOGUE OF CP--VIOLATION}
\end{center}
\vspace{1cm}

\begin{center}
{\large Decio Cocolicchio$^{(1,2)}$ and Luciano Telesca$^{(3)}$}
\end{center}
\begin{center}
$^{(1)}$
{\it
Dipartimento di Matematica, Univ. Basilicata, Potenza, Italy\\
Via N. Sauro 85, 85100 Potenza, Italy}
\end{center}
\begin{center}
$^{(2)}$
{\it
Istituto Nazionale di Fisica Nucleare, Sezione di Milano, Italy\\
Via G. Celoria 16, 20133 Milano, Italy}
\end{center}
\begin{center}
$^{(3)}$
{\it
Consiglio Nazionale delle Ricerche, Istituto di Metodologie 
Avanzate\\
C/da S. Loya TITO, Potenza, Italy}
\end{center}
\vspace{0.5cm}
P.A.C.S. number(s): 
11.30.Er,~~ 	
13.20.Eb,~~	
13.25.+m,~~	
14.40.Aq	
\vspace{1.5cm}
\begin{abstract}
\noindent
The phenomenological features of the mixing in the neutral 
pseudoscalar mesons $\kok$ can be illustrated in the classical 
framework of mechanics and by means of electromagnetic coupled 
circuits. The time-reversed not-invariant processes and the related 
phenomenon of $CP$-nonconservation can be induced by dissipative 
effects which yield a not vanishing imaginary part for the relevant 
Hamiltonian. Thus, two coupled dissipative oscillators can resemble the
peculiar asymmetries which are so common in the realm
of high energy particle physics.
\end{abstract}
\normalsize

\vfill~\vfill~
\thispagestyle{empty}
\newpage
\baselineskip=12pt
\section{The Major Issues of CP Violation}

\bigskip
\noindent
Symmetries are one of the basic cornerstones in the formulation of the 
laws of nature leading to conservative quantities. Unexpected 
violations of symmetries indicate some dynamical mechanisms underlying 
the current understanding of physics. In this context, the space-time
discrete symmetries and their violation represent one of the
most controversial topic. For a long time, the fact that Maxwell 
equations were invariant under space-inversion or parity ($P$) and 
time-reversal ($T$) bolstered the idea that all the laws of physics 
are invariant under those discrete operations. It was easily seen that 
electromagnetic equations possess another discrete symmetry since they 
are unaffected by a charge conjugation ($C$) operation which reverses 
the sign of all the charges and converts a particle into its 
antiparticle. However, since 1957 \cite{WU}\ 
we know that parity is violated in 
weak interactions among fundamental particles.
During early sixties, $CP$ was supposed to be conserved although, $C$ 
and $P$ were individually violated. Since 1964 \cite{CCFT}, we know 
that $CP$ is also violated although to a much lesser extent. 
$CP$ globally stands for the operation that takes all particles into their 
mirror-image antiparticles (and viceversa). If the universe started 
out in a $CP$-symmetric state, and if the laws of physics were 
$CP$-invariant, that is, if $CP$ were conserved, then the world would 
not be in its present asymmetric state, and we would not be here. 
On the 
other hand, the origin of $CP$-violation is still not explained since 
$CP$-violating tiny effects are known to be smaller than the usual 
weak interaction strength by about three orders of magnitude and it
is not excluded that $CP$-violation could be an indication of some effects 
of new physics at a much higher energy scale. The only 
almost overwhelming theoretical prejudice comes against $CPT$ 
violation.
There are very strong reasons \cite{CPT} to believe that fundamental 
interactions can never violate $CPT$ invariance. This again would have 
been wrong \cite{CPTnoft} in some extension of the Lorentz 
invariant and local commutativity field theories,
like in the case of string theory \cite{CPTnostr}.
We must also say that the experimental evidence on $CPT$ is very 
limited at present. It is concerned with the equality of masses and 
lifetimes, and the reversal of magnetic moments between particle and 
antiparticle. Thus, except some loosing theoretical models,
the validity of $CPT$ is assumed and consequently the $T$ violation is 
supposed to be reflected immediately in a $CP$ counterpart.
However, it should be borne in mind that observation of a $T$ odd 
asymmetry or correlation is not necessarily an indication of $CP$
(or even $T$) violation. The reason for this is the anti-unitary 
nature of the time reversal operator in quantum mechanics. As a 
consequence of this, a $T$ operation not only reverses spin and 
three-momenta of all particles, but also interchanges initial and 
final states. Put differently, this means that final-state 
interactions can mimic $T$ violation, but not genuine $CP$ violation.
Presently, genuine experimental evidence of CP--violation comes from the 
mixing and decays of the unstable two-level neutral kaon system. In 
literature exists a large extension of successful modelling and 
phenomenological computations~\cite{Buras}.
Although the discovery of $CP$-violation 
indicated that the kaon system is somewhat more complex
than the typical two-state 
problem and involves considerably more subtle complications;
there are many ways to illustrate this two-state system in
classical physics~\cite{BW}.
From the point of view of Quantum Mechanics,
the $\kak$ represents a complex system consisting 
of two communicating metastable states.
Such puzzle system can be related to the problem of two coupled 
degenerate oscillators with dissipation.
In fact, the intrinsic dissipative nature of this unstable system
and its decay sector, faced with the problem of the complex eigenvalue of
the Hamiltonian and therefore with the extension of the Hilbert space
which is a common tool to deal with open systems far from equilibrium.

\noindent
In particle physics, the neutral kaon decays exhibit unusual and 
peculiar properties arising from the degeneracy of $K^0$ and 
$\kob$. Although $K^0$ and $\kob$ are expected to be distinct 
particles from the point of view of the strong interactions, they 
could transform into each other through the action of the weak 
interactions. In fact, the system $\kok$  results degenerate due to
a coupling to common final states (direct $CP$-violation)
($K^0\leftrightarrow\pi\pi\leftrightarrow {\kob} $) or by means of a 
mixing ($K^0 \leftrightarrow \kob $) (indirect $CP$ 
violation). Then, it is one of the most important tasks of physics
to understand the sources of $CP$-asymmetry~\cite{Coco}.
Even if this is an effect of second order in the weak interactions, 
the transition from the $K^0$ to $\kob$
becomes important because of the degeneracy.
$K^0$ and $\kob$ mesons are charge $C$ conjugate of one another and
are states of definite strangeness $+1$ and $-1$ respectively (conserved
in strong productions $\Delta S=0$, violated in weak decays $\Delta S=1$).
However, they do not have definite lifetimes for weak decay nor do they
have any definite mass, this means that they are not eigenstates.
The mass eigenstates are linear combinations
of the states $\kk$  and $\kbk$ namely $\kks$ and $\kkl$,
which have definite masses and lifetimes.
The short lived $\kks$ meson decays with a characteristic time
$\tau_S \ = \ (0.8922 \ \pm \ 0.0020)\ {10}^{-10} \ sec$
into the two predominant modes $\pi^+
\pi^-$ and $\pi^0\pi^0$ each with the CP eigenvalue $+1$, whereas the long
lived $\kkl$ mesons with a decay time
$\tau_L \ = \ (5.17 \ \pm \ 0.04)\ {10}^{-8} \ sec$
has among its decay modes $\pi^+\pi^-\pi^0$ and
$\pi^0\pi^0\pi^0$, which are eigenstates of CP with eigenvalue $-1$.

In 1964, it was observed~\cite{CCFT}
that there is a small but finite probability for the decay
$K^0_L\rightarrow\pi^+\pi^-$, in which the final state has the CP eigenvalue +1.
Thus we cannot identify $K^0_L$ with $K^0_1$ and $K^0_S$ with $K^0_2$, 
the real eigenstates of $CP$-symmetry.

\noindent
Let $\kos$ and $\kobs$ be the stationary states of the
$\ko$-meson and its antiparticle $\kob$, respectively. Both
states are eigenstates of the strong and electromagnetic interactions.
The $\ko$ and $\kob$ states are connected through
CP transformations. With the exception of an arbitrary phase
we obtain
\beq
\begin{array}{c}
C\!P \, \kos = e^{i \, \theta _{\rm CP} } \kobs~~{\rm and}~~
C\!P \, \kobs = e^{- \,i \, \theta_{\rm CP} } \kos~~ \\
T \, \kos = e^{ i \, \theta_{\rm T} } \kos~~{\rm and}~~
T \, \kobs = e^{ i \, \overline {\theta}{}_{\rm T } } \kobs
\end{array} \label{cp-phase}\eeq
where $\theta$'s are arbitrary phases and it follows that
\beq
2 \, \theta_{\rm CP} =\overline{\theta}{}_{\rm T}\, - \, \theta_{\rm
T}~.
\eeq
by assuming $C\!PT\, \kos = TC\!P \, \kos$.
As stated before, $\kos$ and $\kobs$ are eigenstates of strange 
flavour.
Since strangeness quantum numbers is conserved in strong interactions,
their relative phase can never be measured. So every observable is 
independent of the phase transformation $\kos\rightarrow 
e^{i\xi}\kos$.
In presence of a new interaction
violating strangeness conservation,
the K-mesons can decay into final states with no strangeness ($|\Delta
S|= 1$) and $\ko$ and $\kob$ can oscillate to each other
($|\Delta S| = 2$). A convenient way to discuss this mixing new
interaction, which is very much weaker than strong and
electromagnetic interactions, require the detailed
knowledge of the perturbative effects \cite{Buras}.
Although this kind of problems require the 
formalism of density matrix, and the notion of
rest frame for an unstable degenerate system
appears difficult to be implemented, nevertheless perturbation theory
and the single pole approximation of Weisskopf-Wigner method
\cite{wiger-weis} can be applied to derive 
the eigenstates  $K_S$ and $K_L$
of a $2 \times 2$ effective Hamiltonian 
${\cal H}$ \cite{Buras}. 
The time evolution of basis states $\k$ and $\bk$ can be
written as
\beq
i{d\over {dt}}|\Psi(t)>={\cal H}|\Psi(t)>=
\left(
\matrix {H_{11} & H_{12} \cr
H_{21} & H_{22} \cr}
\right)
|\Psi(t)>
\eeq
\cp
where the effective Hamiltonian ${\cal H}$ 
is an arbitrary matrix 
which can be 
uniquely rewritten in terms of two Hermitian matrices
$M$ and $\Gamma$: ${\cal H} = M - i \Gamma /2$.
\cp
M is called the mass matrix and $\Gamma$ the decay matrix.
Their explicit expressions can be derived by the weak scattering matrix
responsible of the decay ${S^w}_{\alpha ', \alpha }$
\beq
{S^w}_{\alpha ', \alpha }=<\alpha '|Te^{-i  \int{H_w'(t)dt}}|\alpha >
\simeq \delta_{\alpha' ,\alpha }+(2\pi )^4 \delta ^4(p-
p')iT_{\alpha' ,\alpha }(p)
\eeq
\noindent
where
$H_w'=e^{iHt}H_we^{-iHt}$ and $H_w$ representing the weak component
of the Hamiltonian, whereas its strong sector gives at rest
\beq
<\alpha '|H_{strong}|\alpha>=m_0\delta_{\alpha ,\alpha '}
\eeq
Therefore we can resume that the matrix elements are given by
\beq
<\alpha '|{\cal H}|\alpha>={\cal H}_{\alpha ',\alpha }=m_0\delta_{\alpha 
,\alpha '}-T_{\alpha ',\alpha }(m_0)
\eeq
\cp
and by means of the following 
relations
\beq
\Theta ={{-1}\over {2\pi i}} \int{ d\omega {{e^{-i\omega t}}\over
{\omega + i\eta}}}
\ \ \ \ \ \ \ \ \ 
{1 \over {\chi - a + i\eta}}=P {1 \over {\chi - a}} - i\pi \delta (
\chi -a)
\eeq
let us derive straightforward that
\beq
\Gamma_{\alpha ',\alpha}=2\pi\sum_n{<\alpha'
|H_w|n><n|H_w|\alpha >\delta(E_n-m_0)}
\eeq
\cp
\beq
M_{\alpha ',\alpha}=m_0\delta_{\alpha ',\alpha}+
<\alpha '|H_w|\alpha >+\sum_n {P{{<\alpha '|H_w|n><n|H_w|\alpha>}\over 
{m_0-E_n}}}
\eeq
\cp
Note that this last sum is taken over all
possible intermediate states whereas for the width we take only real 
final states common to $\ko$ and $\kob$.
\cp
It can be shown that if CPT holds then the restriction ${\cal H}_{11} 
= {\cal H}_{22}$ and hence $M_{11} = M_{22}, ~ \Gamma_{11} =
\Gamma_{22}$ must be adopted. 
Furthermore if $CP$ invariance holds too, then besides
$H_{11}=H_{22}$ we get also $H_{12}=H_{21}$ and consequently
$\Gamma_{12}=\Gamma_{21}={\Gamma_{21}}^*$, $M_{12}=M_{21}={M_{21}}^*$
so that $\Gamma_{ij}$, $M_{ij}$ will result all real numbers.
\cp
Since, $\ko$ e $\kbo$ are only the eigenstates of the strong 
Hamiltonian, the eigenvalues of the effective Hamiltonian ${\cal H}$
can be found by means the diagonalization which yields
\beq
\lambda_S = H_{11} -\sqrt {H_{12}H_{21}}=M_{11}-{i\over 2} \Gamma_{11}
-Q=m_S -{i\over 2} \Gamma_S
\eeq

\beq
\lambda_L = H_{11} +\sqrt {H_{12}H_{21}}=M_{11}-{i\over 2} \Gamma_{11}
+Q=m_L -{i\over 2} \Gamma_L 
\eeq
\cp
where
\beq
Q=\sqrt{ H_{12} H_{21} } = 
\sqrt{(M_{12}-{i\over 2} \Gamma_{12}) ({M_{12}}^*-{i\over 2} 
{\Gamma_{12}}^*)} .
\eeq
These real (m) and imaginary ($\Gamma$) components will define the 
masses and the decay width of the ${\cal H}$ eigenstates $K_S$ e $K_L$. 
\noindent
The mass difference and the relative decay width are easily given by
\beq
\Delta m=m_L-m_S=2Re Q\simeq 2 \hbox{Re} {M_{12}}= 
(0.5351\pm0.0024)\cdot 10^{-10}\hbar sec^{-1}
\eeq

\beq
\Delta \Gamma =\Gamma_S - \Gamma_L = 4 ImQ\simeq 2 \hbox{Re} 
\Gamma_{12} =  (1.1189\pm 0.0025)\cdot 10^{-10} sec^{-1}
\eeq

The experimental evidence \cite{CCFT} in 1964 that both the short-lived 
$K_S$ and long-lived $K_L$ states decayed to $\pi \pi$ upset 
this tidy picture.
It means that the states of definite mass and lifetime are never 
more states with a definite $CP$ character.
\noindent
With the conventional choice of phase, the $CP$ eigenstates $K_1$ and 
$K_2$ enter into play so that the mass eigenstates
can be parameterized by means of an impurity complex parameter
$\epsilon$ which encodes the indirect mixing effect of CP violation 
in the neutral kaon system.
The general expression of the relative eigenstates can be 
obtained by the standard procedure of diagonalization. 
\beq
|K_S>= {1\over \sqrt{2(1+|\epsilon_S |^2)}} \left [ (1+ \epsilon_S )|K^0> +
(1- \epsilon_S )|\overline K^0>\right ]= 
{|K_1> + \epsilon_S |K_2>\over \sqrt {1+|\epsilon_S |^2}}
\eeq

\beq
|K_L>= {1\over \sqrt{2(1+|\epsilon_L |^2)}} \left [ (1+ \epsilon_L )|K^0> 
-(1- \epsilon_L )|\overline K^0>\right ]= 
{|K_2> + \epsilon_L |K_1>\over \sqrt {1+|\epsilon_L |^2}}
\eeq
\cp
where $\epsilon_S=\epsilon +\Delta$, $\epsilon_L=\epsilon -\Delta$, being
\beq
\epsilon={{[-ImM_{12}+iIm{{\Gamma_{12}} \over 2}]}\over {{{(\Gamma_L-
\Gamma_S)}\over 2}+i(m_L-m_S)}}
\qquad
\Delta={1\over 2}{{i[(M_{11}-M_{22})+{{(\Gamma_{11}-\Gamma_{22}}\over 
2})]}\over {{{(\Gamma_L-\Gamma_S)}\over 2}+i(m_L-m_S)}}
\eeq
The $\Delta$ parameter vanishes if we assume CPT invariance.
Presently, the experimental data \cite{PDG}\  are quite consistent with the 
expectation of the spin--statistics connection which excludes CPT 
violation in any local, Lorentz invariant theory. 
So that,
we can suppose that $\Delta =0$ and consequently $\epsilon_L = 
\epsilon_S$. 
At present, the magnitude is 
$|\epsilon| \ = \ (2.259 \ \pm \ 0.018)\ {10}^{-3} $
and its phase results $\Phi_{\epsilon} \ = \ {(43.67 \ \pm \ 0.13)}^o $.
Notice that, in contrast to $|K_1
\rangle$ and $|K_2 \rangle$, the states $|K_S \rangle$ and $|K_L \rangle$
are not orthogonal to one another but have a not vanishing scalar product
\beq
<K_S|K_L>= {{2 Re\epsilon
+i Im \Delta} \over {1+|\epsilon |^2}}
\eeq
Obviously, if CP were conserved, $\epsilon$ would vanish or would 
reduce to a pure phase, which can be reabsorbed in the redefinition 
of $\ko$, $\kbo$ and $K_S=K_1$, $K_L=K_2$ would result.
A brief comment is required in previous formula where we propose as the
impurity parameter a term $\epsilon$ which does not take into
account the direct contribution to the mixing $\kok$
due to the coupling via the common $\pi\pi$ decay sector.
Such a model, called superweak, can likely be selected in the next 
future \cite{Coco}.

The small amount of the $CP$ violation is indeed reflected in the 
impurity parameter
\beq
\epsilon={e^{i{\pi \over 4}}\over {2\sqrt {2}}} \left( {{ImM_{12}}\over 
{ReM_{12}}}\ -{i\over 2}{{Im\Gamma_{12}}\over 
{ReM_{12}}} \right)=
{e^{i{\pi \over 4}}\over {\sqrt{2} \Delta m}} \left( { ImM_{12} + 2 
\xi_0 \hbox{Re} M_{12} } \right) 
\eeq
where $\xi_0$ represents the additional manifestations in the mixing of 
the $CP$ violation due to the effective isospin decay amplitudes.
In order to derive the amount of this complex parameter, it appears 
more convenient to write the eigenstates of ${\cal H}$ as
\beq\eqalign{
|K_L \rangle =& p | \k \rangle + q | \bk \rangle =
{1\over {\sqrt{1+\vert\alpha\vert^2}}} (\kk+\alpha \kbk)\; ,\cr
|K_S \rangle =& p | \k \rangle - q | \bk \rangle =
{1\over {\sqrt{1+\vert\alpha\vert^2}}} (\kk-\alpha \kbk)\; ,\cr } 
\eeq
where $|p|^2 + |q|^2 = 1$. With a proper phase choice, we introduced 
the new relevant parameter
\beq
\alpha = {q \over p} = {{1 - \epsilon}\over {1 + \epsilon}}=
\sqrt{\frac{{\cal H}_{21}}{{\cal H}_{12}}} = \sqrt{
{ {M_{12}^* - {i\over 2}\Gamma_{12}^*}\over
  {M_{12} - {i\over 2}\Gamma_{12} } } }
\eeq
and then
\beq \label{eqn:eps}
\eqalign{
\epsilon =& \frac{p-q}{p+q} = \frac{1-\alpha}{1+\alpha}=
\frac{\sqrt{\h_{12}} - \sqrt{\h_{21}}}{\sqrt{\h_{12}} + 
\sqrt{\h_{21}}}=\cr
&= { {2i \hbox{Im}M_{12} + \hbox{Im}M_{12}} \over 
{ (2\hbox{Re}M_{12} - i \hbox{Re}\Gamma_{12}) + 
(\Delta m -{i\over 2} \Delta\Gamma) } }
\simeq
\frac{i~{\rm Im} M_{12} + {\rm Im} (\Gamma_{12}/2)}{(\Delta m 
- {i\over 2} \Delta\Gamma)} \cr } 
\eeq
Thus it results evident that the CP-violation parameter $\epsilon$
arises from a relative imaginary part between the
off-diagonal elements $M_{12}$ and $\Gamma_{12}$ i.e. if 
$arg(M_{12}^*\Gamma_{12})\ne 0$.
If we rewrite in polar coordinate the complex ratio between these 
relevant components
\beq
{{M_{12}}\over{\Gamma_{12}}}=-
{ {\vert{M_{12}\vert} \over {\vert{\Gamma_{12}}\vert} } }e^{i\delta}
= r e^{i \delta}
\eeq
where due to the smallness of $\vert \epsilon\vert$
($\hbox{Im}M_{12}\leq\hbox{Re}M_{12}$ and
$\hbox{Im}\Gamma_{12}<<\hbox{Re}\Gamma_{12}$)
\beq
r\simeq {{\Delta m}\over{\Delta \Gamma}}=(0.477\pm 0.003)
\eeq
It is clear that only the parameter $|\alpha|$ is significative.
In the sense that $\alpha\ne 1$ does not necessarily imply $CP$ 
violation. $CP$ is violated in the mixing matrix if $|\alpha|\ne 1$.
Remember that, since flavour is conserved in strong interactions, 
there is some freedom in defining the phases of flavour eigenstates. 
This means that $\alpha= q/p$ is a phase dependent quantity 
manifesting its presence in the phase of $\epsilon$ which must only 
satisfy
\beq
<K_S|K_L>= x = {{2 Re\epsilon} \over {1+|\epsilon |^2}}
\eeq
which reduces to the equation of a circle
\beq
(\hbox{Re}\epsilon - {1\over x})^2 +(\hbox{Im}\epsilon)^2 = ({1\over 
x})^2 -1
\eeq
of radius $\sqrt{({1\over x})^2 -1}\simeq {1\over x}$ centered in 
($1\over x$, 0) in the Gauss complex $\epsilon$-plane.

The relative pure phase $\delta$ can be derived from the fact that
\beq
\alpha = {q \over p} = {{1 - \epsilon}\over {1 + \epsilon}}\simeq
1 - { {2r}\over {4r^2 +1}} ( 1+ 2 i r)\delta
\eeq
The amount of its value can then be extracted only by analyzing the 
experimental results of the semileptonic asymmetry
\beq
A_{SL} =
{{\Gamma(K_L\rightarrow \ell^+\nu X)-\Gamma(K_L\rightarrow \ell^-\nu X)}
\over
{\Gamma(K_L\rightarrow \ell^+\nu X) +\Gamma(K_L\rightarrow \ell^-\nu X)}}
=
{{1-|\alpha|^2}\over{1+|\alpha|^2}}
\simeq { {2r}\over{4r^2 +1}}\delta .
\eeq
Its experimental value is $A_{SL} = (3.27\pm 0.12)\cdot 10^{-3}$
and then the relative phase $\delta=(6.53\pm 0.24)\cdot 10^{-3}$.
In the Wu-Yang convention $\hbox{Im}\Gamma_{12}=0$, we obtain that
\beq
\arg(\epsilon)\simeq\cases{
\pi -\Phi_{SW}\quad\hbox{for}\quad\hbox{Im}{M_{12}}>0\cr
\Phi_{SW}\quad\hbox{for}\quad\hbox{Im}{M_{12}} < 0\cr}
\eeq
being the superweak phase $\Phi_{SW}=\tan^{-1}(2 r)$.
The matrices $\Gamma$ and $M$ may be expressed perturbatively
\cite{Buras} in terms of sums
over states connected to $\k$ and $\kbo$ by means of the weak Hamiltonian $H_W$.
By considering specific electroweak models for
the kaon decays into $2 \pi,~3 \pi,~\pi l \nu$, and other final states, one
can then compare theory and experiments.

In general, we can resume that 
there are three complementary ways to describe the evolution of the
complex neutral kaon system:
\medskip

{1)}{  In terms of the mass eigenstates $K_{L,S}$, which do not posses
definite strangeness
\beq
\eqalign{
\kst=&\kso\eas\qquad\alpha_S={\it i}m_S+{{\gamma_S}\over 2}\cr
\klt=&\klo\eal\qquad\alpha_L={\it i}m_L+{{\gamma_L}\over 2}\cr}
\eeq
\smallskip

{2)}{  In terms of the flavour eigenstates, whose time evolution
is more complex
\beq
\eqalign{
\kot=&\fpt\koo+\alpha\fmt\bkoo\cr
\bkot=&{1\over\alpha}\fmt\koo+\fpt\bkoo\cr}
\eeq
with
\beq
\eqalign{
\fpt=&{1\over 2}(\eas+\eal)={1\over 2}\eas\left[1+e^{-(\alpha_L-\alpha_S)t}\right]\cr
\fmt=&{1\over 2}(\eas-\eal)={1\over 2}\eas\left[1-e^{-(\alpha_L-\alpha_S)t}
\right]\quad.
\cr}
\eeq
{3)}{ In terms of the $CP$--eigenstates $K_1$ and $K_2$  
\beq
\eqalign{
\kku=&{1\over{\sqrt 2}}(\kk+\kbk)\cr
\kkd=&{1\over{\sqrt 2}}(\kk-\kbk)\cr}\qquad\qquad
\eqalign{
CP\kku=&+\kku\cr
CP\kkd=&-\kkd\cr}
\eeq
which we let us express the mass eigenstates as 
\beq
\eqalign{
\kks=&{1\over{\sqrt 2}}\left[ (p+q) \kku+(p-q)\kkd)\right]\cr
\kkl=&{1\over{\sqrt 2}}\left[ (p-q) \kku+(p+q)\kkd)\right]\cr}
\eeq
\medskip
The three bases $\{K_S,K_L\}$, $\{K^0,{\overline K}^0\}$ 
and $\{ K_1$, $K_2\} $, are completely equivalent.

\section{\bf The Mechanical Analogue of the $\kok$
Complex System.}
\bigskip
We seek in classical physics an analogue of the two-state mixing problem which
leads to a non-zero value of $\epsilon$. 
The problem is quite difficult since the equations of motion in 
classical mechanics are time-reversal invariant. The main features of 
irreversibility enter only considering the effects of dissipation.
Anyway, these results seem to reflect the well known requirements
of additional complementarity relations,
which occur at classical level,
to make the equations of motion of dissipative systems derivable from 
a variational principle \cite{Bate}.
The neutral kaon system resembles just the typical problems we meet in
coupled degenerate oscillators with dissipation in classical physics.  
A system of two coupled pendula~\cite{BW} provides the simplest analogy to the
two level system.  The requirement of CPT invariance is satisfied by taking
the two pendula to have equal natural frequencies $\omega_0$ (and, for
simplicity, equal lengths and masses).  If one couples them by a connecting
spring, the two normal modes will consist of one with frequency $\omega_1 =
\omega_0$ in which the two pendula oscillate in phase, and another with
$\omega_2 > \omega_0$ in which the two oscillate 180$^{\circ}$ out of phase,
thereby compressing and stretching the connecting spring.  If the connection
dissipates energy, the mode with frequency $\omega_2$ will eventually decay
away, leaving only the in-phase mode with frequency $\omega_1 = \omega_0$.

\noindent
Another mechanical analogue of the two-level system can be obtained
using two coupled oscillators, each consisting of a identical mass $m$ and
spring constant $k_1$ coupled by a $k_2$ spring constant.
The way to obtain the equations of their motion is straightforward. 
Supposing the mass unitary we get that
the equation of the motion of the two coupled oscillators are
\beq
\ddot x_1 = -k_1 x_1 - k_2 ( x_1 -x_2 )~~~
\eeq
\beq
\ddot x_2 = -k_1 x_2 + k_2 ( x_1 -x_2 )~~~
\eeq
\noindent
Now we assume an harmonic behavior:
$x_i = X_i e^{-i \omega t}$, and solve the two coupled equations in
$x_1$ and $x_2$ for characteristic values of $\omega$:
\beq 
( k_1 +k_2 - \omega^2  )^2 = k_2^2~~~.
\eeq
\noindent
One solution has the natural frequency $\omega^2 = k_1$ independently of the
coupling.  This is the solution with $x_1 = x_2$, the symmetric mode. 
The other solution has a frequency $\omega^2 = k_1 + 2 k_2$, which corresponds
to the asymmetric mode, $x_1 \ne x_2$. The spring constant $k_2$ produces a
frequency difference between the two normal modes.
The symmetric solution with $x_1 = x_2$ corresponds to a symmetric 
whereas the asymmetric solution corresponds to $x_1 \ne x_2$.
We consider, now, the effects of dissipation for the
two-state oscillator system introducing a small dissipative
coupling
between the two masses in the form of
an air dash-pot (Fig. 1).
This device gives a velocity-dependent  force of the form
\beq
f= -k \dot x .
\eeq
In the particular case of the two coupled oscillators the dissipation force on
a mass can be written as a velocity-dependent force due to the other
mass. 
The equations of motion contain a dissipative term and can be 
expressed as
\beq 
\ddot x_1 = -( k_1 + k_2 ) x_1 +k_2 x_2 + a \dot x_2
\eeq
\beq 
\ddot x_2 = -( k_1 + k_2 ) x_2 +k_2 x_1 - a \dot x_1
\eeq
We can rewrite these equations by means of the standard formalism of 
analytical mechanics
\beq
\frac{d}{dt} \frac{\partial T}{\partial x_j} -\frac{\partial T}{\partial t}
= Q_j
\eeq
where
\beq
T= \frac{1}{2} \sum_{i,j}\delta_{ij} \frac{d x_i}{dt}\frac{dx_j}{dt}\; ,
\eeq
and of the generalized momenta
\beq
Q_j =  \sum_{i=1}^{2} A_{i j} x_{i} + B_{i j} \frac{dx_i}{dt}\; .
\eeq
being
\beq
\eqalign{
A_{11}&= A_{22} = - (k_1 +k_2)\cr
A_{12}&= A_{21} = k_2\cr
B_{11}&= B_{22} = 0\cr
-B_{12}&= B_{21}=a\cr}
\eeq

A normal mode solution of the kind $x_i = X_i e^{-i \omega t}$
can be derived from the secular equation

\beq
\vert \delta_{ij} \omega^2 + A_{ij} + i \omega B_{ij}\vert=0
\eeq
In order to recast the mechanical problem in a form closer to previous
relations, we rearrange the problem in a complex vectorial space of 
constituent base $\ket{\eu}$ and $\ket{\ed}$. So that any vector state
\beq {\boldmath x} \equiv [ \begin{array}{c} x_1 \\ x_2 \end{array}
]~~~.
\eeq
can be expressed as
\beq
\ket{{\boldmath x}} = x_1 \ket{\eu} + x_2 \ket{\ed}.
\eeq
Thus the equations of the motion can be rewritten as
\beq
{\cal I} \frac{d^2 | x >}{dt^2} = {\cal K} | x > 
- {\cal A} \frac{d| x >}{dt}
\eeq
with
\beq
{\cal K} \equiv [ \begin{array}{c c}
-( k_1 + k_2 )& k_2  \\
 k_2  & -(k_1 +k_2 ) \\
\end{array} ]~~~,
\eeq
 
\beq
{\cal A} \equiv [ \begin{array}{c c}
0   & -a  \\
a  & 0 \\
\end{array} ]~~~,
\eeq

Its physical solution derive from the characteristic equation
\beq
{\cal H} {| \bf x >} = \omega^2 {| \bf x >}~~~,
\eeq
being
\beq
{\cal H} = i \omega {\cal A} - {\cal K} =
[ \begin{array}{c c}
 k_1 + k_2 & -k_2 -i \sqrt{k_1} a \\
 -k_2 +i \sqrt{k_1} a & k_1 +k_2 \\
\end{array} ]~~~,
\eeq
Here we have replaced $\sqrt{k_1}$ with $\omega$ in the coupling term.

The violation of ``CP''  invariance is  parameterized by
the presence of small antisymmetric off-diagonal terms in ${\cal H}$.

The characteristic equation now becomes
\beq
(k_1 - \omega^2)^2 + 2 k_2 (k_1- \omega^2) - a^2 k_1= 0~~~.
\eeq

The physical solutions are normal modes with $\omega^2 \simeq k_1 +2 k_2$,
corresponding to $K_S$ (whose mass is affected strongly by the
``CP-conserving'' coupling $k_2$),
and with $\omega^2 = k_1 - \frac{a^2 k_1}{2 k_2}$, corresponding
to $K_L$ (whose mass now also receives a small contribution from the
``CP-violating'' coupling $a$).  Thus the  general solution
recovers the notation of the kaon system 
with the identification
\beq
\epsilon \simeq \frac{-i a  \sqrt {k_1}}{2 k_2}~~~.
\eeq
and
\beq
\frac{q}{p} = \sqrt{ \frac{-k_2 + i a  \sqrt {k_1}}
{-k_2 - i a  \sqrt {k_1}}}~~~.
\eeq

Here we have assumed $a$ sufficiently small so that $|\epsilon| \ll 1$.
The relative strength and the phase of the antisymmetric and symmetric couplings
thus governs $\epsilon$, just like in the case of neutral kaons.
\bigskip

\section{\bf The Electrical Analogue of CP Violation}
\bigskip

The dynamics of two state kaon system can be also reproduced by a 
generalization of engineer's transmission line in load problems and, 
as such, by means of any problem with degenerate constituents which 
involve standing wave and dissipation.
A simple electrical analogue (Fig. 2) can be built up
using two $L-C$ ``tank'' circuits, each consisting of a capacitor $C_i$ in
parallel with an inductor $L_i$ and having resonant frequency $\omega_{0i} =
(L_iC_i)^{-1/2}$ $(i = 1,2)$. The two tank circuits are coupled through a
two-port device $G$, so that the voltages  and the currents on each circuit
are related to each other by a linear relation, that is representative of the
2-port device, 

\beq
\eqalign{
f_1(I_1,I_2 ; V_1,V_2)&=0\cr
f_2(I_1,I_2 ; V_1,V_2)&=0\; ,\cr}
\eeq
with $f_1$ and $f_2$ linear functions.
A 2-port component can have almost six representation, that are 2x2 matrices.
If we designate as $h_1$ and $h_2$ two of the four electrical variables 
$I_1$, $I_2$,
$V_1$, $V_2$, and as $h^\prime_1$ and $h^\prime_2$
the remaining two variables, we can write a 
matrix relation 
\beq
{\cal G}\, {\it h_{inp}} = {\it h_{out}}~~~,
\eeq
where
\beq
{\cal G} \equiv \left[ \begin{array}{c c}
g_{11} & g_{12} \\
g_{21} & g_{22} \\
\end{array} \right]~~~,
\quad\hbox{where}\quad 
{\it h_{inp}} \equiv \left[ \begin{array}{c} h_1 \\ h_2 \end{array}
\right]~~~
\qquad\hbox{and}\;\;
{\it h_{out}} \equiv \left[ \begin{array}{c} h^\prime_1 \\ 
h^\prime_2 \end{array}
\right]~~~.
\eeq
If we choose $h_1=V_1$, $ h_2=V_2$, and $h^\prime_1=I_1$,  
$h^\prime_2=I_2$,
whereas $G$ represents the admittance matrix.
The equations of the circuit become:
\beq
\eqalign{
I_1&= g_{11} V_1 + g_{12} V_2\cr
I_2&= g_{21} V_1 + g_{22} V_2\; ,\cr}
\eeq
If we assume, for simplicity,  $L_1 = L_2 = L$ and $C_1 = C_2 = C$, 
and we calculate the currents
flowing in the two tank circuits, the symbolic solutions
$V_i = v_i e^{-i \omega t}$, $I_i =  i_i e^{-i \omega t}$
give the voltages $v_i$:
\beq
 v_i = \frac{i \omega L}{1- \omega^2 L C} i_i \quad .
\eeq
Substituting in the equation of the 2-port device, it is straightforward 
to deduce the following matrix equation:
\beq
{\cal G'} {\bf i} = \omega^2 {\bf i}~~~,
\eeq
where
\beq
{\cal G'} \equiv \left[ \begin{array}{c c}
\omega_0 ^2 - \frac{i g_{11} \omega_0}{C} & \frac{-i g_{12} \omega_0 }{C} \\
\frac{-i g_{21} \omega_0}{C} & \omega_0 ^2 - \frac{i g_{22} \omega_0}{C} \\
\end{array} \right]~~~,
\eeq
and we have replaced $\omega$ by $\omega_0$ in the matrix representation.

We see that CP-violation can be obtained if $g_{12} \ne g_{21}$
(this condition is usually said not-reciprocal).
If we take $g_{11} \simeq g_{22}$ , $g_{12} \simeq g_{11} - g^* $ and 
$g_{12} \simeq g_{11} + g^*$, with small antisymmetric off-diagonal terms 
$g^*$, we can easily recognize that the solution  
$\omega^2 = \omega_0 ^2 - 2 i g_{11} \frac{\omega_0}{C}$ 
corresponds to $K_S$ if
\beq
\epsilon \simeq \frac{g^*}{2 g_{11}}~~~.
\eeq
\vfill

\bigskip
\centerline{\bf ACKNOWLEDGMENTS}
\bigskip
One of us (L. T.) wishes to thank the warm hospitality of
the Institute of Advanced Methodologies of Environmental
Analysis and gratefully acknowledges the financial 
support of a grant of the National 
Research Council (Consiglio Nazionale delle Ricerche).


\bigskip


\begin{thebibliography}{99}
\bibitem{WU} C. S. Wu, E. Ambler, R. W. Hayward, D. D. Hoppos and R. 
P. Hudson, "{\it Experimental Test of Parity Conservation in Beta 
Decay}", Phys. Rev. {\bf 105} (1957) 1413.

\bibitem{CCFT} J. H. Christenson, J. W. Cronin, V. L. Fitch, and R. Turlay,
"{\it Evidence for the $2 \pi$ decay of the $K_2^0$ meson}",
Phys. Rev. Lett. {\bf 13} (1964) 138.

\bibitem{CPT} R. F. Streater and A. S. Wightman,
"{\it CPT, Spin, Statistics and All That}" (Benjamin, New York, 1964);
N. N. Bogolubov, A. A. Logunov and I. T. Todorov,
"{\it Introduction to Axiomatic Quantum Field Theory}" 
(Benjamin, New York, 1975).

\bibitem{CPTnoft}
A. I. Oksak and I. T. Todorov, 
"{\it Invalidity of TCP-Theorem for Infinite Component Fields}",
Commun. Math. Phys. {\bf 11} (1968) 125.

\bibitem{CPTnostr}
E. Witten, "{\it Elliptic Genera and Quantum Field Theory}",
Commun. Math. Phys. {\bf 109} (1987) 525;
H. Sonoda, "{\it Hermiticity and CPT in String Theory}", \npb{326}{1989}{135};
V. Kostelecky and A. Potting, "{\it CPT and Strings}", \npb{359}{1991}{545}.

\bibitem{Buras} 
See, e.g., P. K. Kabir, {\it The CP Puzzle} (Academic Press,
New York, 1968);
R. G. Sachs, {\it The Physics of Time Reversal
Invariance} (University of Chicago Press, Chicago, 1988);
T. P. Cheng and L. F. Li, {\it Gauge Theory of Elementary
Particles} (Oxford University Press, 1984);
C. Jarlskog, {\it CP--Violation} (World Scientific, 
Singapore, 1989); A. J. Buras, "{\it CP Violation: Present and 
Future}", Proceedings 1st Int. Conf. on Phenomenology of Unification, 
Rome, 1994.

\bibitem{BW} B. Winstein, "{\it CP violation}", in {\it Festi-Val -- Festschrift
for Val Telegdi}, ed.~by K. Winter (Elsevier, Amsterdam, 1988), 
pp.~245-265; L. Telesca, "{\it L'analogo classico della violazione di 
$CP$}", preprint Univ. Basilicata (in Italian) (1994); 
J. Rosner,"{\it Table-top Time-Reversal Violation}"
preprint Univ. Chicago, EFI-95-51.

\bibitem{Coco} D. Cocolicchio and L. Maiani,
"{\it Lepton Asymmetry in B Decays in the Superweak and in the Standard Theory
of CP Violation}", {Physics Letters} {\bf B291} (1992) 155.

\bibitem{wiger-weis} V. Weisskopf and E. Wigner, 
"{\it Berechnung der nat\"urlichen Linienbreite auf Grund der Diracschen 
Lichttheorie}", Z.\ Phys.\ {\bf 63}, 54 (1930)
T. D. Lee, R. Oehme and C. N. Yang, "{\it Remarks on possible
noninvariance under time reversal and charge conjugation}",
Phys. Rev. {\bf 106} (1957) 340.

\bibitem{PDG} Particle Data Group, L. Montanet \ite, "{\it Review of Particle
Properties}", Phys. Rev. {\bf 50} (1994) 1173.

\bibitem{Bate} H. Bateman, "{\it On Dissipative Systems and Related 
Variational Principles}", Physical Review {\bf 38} (1931) 815;
H. Dekker, "{\it Classical and Quantum Mechanics of the Damped 
Oscillator}", Phys. Rept. {\bf 80} (1980) 1.

\end{thebibliography}
\end{document}